\documentclass{emulateapj}
\usepackage{psfig}
\usepackage{apjfonts}

\received{} \revised{} \accepted{}

\shorttitle{Local Ionized X-ray Absorbers} \shortauthors{Fang et al.}

\makeatletter

\newenvironment{figurehere}
   {\def\@captype{figure}}
   {}
\makeatother

\begin{document}

\newcommand{\cms}{$\rm cm^{-2}\ $}
\newcommand{\cmq}{$\rm cm^{-3}$}
\newcommand{\emo}{{\rm EM}_{\rm obs}}
\newcommand{\kms}{$\rm km\ s^{-1}$}
\newcommand{\lpc}{L_{\rm pc}}
\newcommand{\caln}{{\cal N}}
\newcommand{\beq}{\begin{equation}}
\newcommand{\eeq}{\end{equation}}
\def\ew		{{\rm EW}}
\def\ewt	{{\rm EW_{\rm th}}}
\def\caln	{{\cal N}}
\def\aa         {{\AA\ }}

\title{A Galactic Origin for the Local Ionized X-ray Absorbers}

\author{Taotao~Fang\altaffilmark{1}, Christopher F. McKee\altaffilmark{1,2},
  Claude~R.~Canizares\altaffilmark{3}, and Mark
  Wolfire\altaffilmark{4}}
\altaffiltext{1}{Department of Astronomy,
  University of California, Berkeley CA 94720,
  fangt@astro.berkeley.edu; {\sl Chandra} Fellow} 
\altaffiltext{2} {Department of Physics, University
  of California, Berkeley CA 94720}
\altaffiltext{3}{Department of Physics and
  Center for Space Research, MIT, 77 Mass. Ave., Cambridge MA 02139}
  \altaffiltext{4}{Department of Astronomy, University of Maryland
  College Park, MD 20742} 

\begin{abstract}

Recent {\sl Chandra} and {\sl XMM} observations of distant quasars
have shown strong local ($z\sim 0$) X-ray absorption lines from highly
ionized gas, primarily He-like oxygen. The nature of these X-ray
absorbers, i.e., whether they are part of the hot gas associated with
the Milky Way or part of the intragroup medium in the Local Group,
remains a puzzle due to the uncertainties in the distance.  We present
in this paper a survey of 20 AGNs with {\sl Chandra} and {\sl XMM}
archival data. About 40\% of the targets show local \ion{O}{7} He\
$\alpha$ absorption with column densities around $10^{16}\ \rm
cm^{-2}$; in particular, \ion{O}{7} absorption is present in all the
high quality spectra. We estimate that the sky covering fraction of
this \ion{O}{7}-absorbing gas is at least 63\%, at 90\% confidence,
and likely to be unity given enough high-quality spectra. Based on (1) the expected number of
absorbers along sight lines toward distant AGNs, (2) joint analysis
with X-ray emission measurements, and (3) mass estimation, we argue
that the observed X-ray absorbers are part of the hot gas associated
with our Galaxy. Future observations will significantly improve our
understanding of the covering fraction and provide robust test of this
result. 

\end{abstract}


\section{Introduction}

Recently, a number of {\sl Chandra} and {\sl XMM}-Newton
observations of quasars have shown local ($z \approx 0$) X-ray
absorption lines
(\citealp{kas02,nic02,fan02,ras03,fan03,cag04,mck04,wil05}). These
background quasars are among the brightest extragalactic X-ray
sources in the sky and some of them were used as calibration
targets. The typical high ionic column densities of these X-ray
absorbers ($\sim 10^{16}$ \cms) imply the existence of large
amounts of hot gas with temperatures around $10^6$ K. Recent ultraviolet observations of the local high velocity \ion{O}{6}
absorbers also reveal such hot gas but at lower temperatures
(see. e.g., \citealp{sem03,nic03}). Given the spectral resolution of {\sl Chandra}
and {\sl XMM}-Newton, it is still unclear where this hot gas is
located: in the interstellar medium, in the Galactic halo, or in
the Local Group as the intragroup medium.   

In sharp contrast, so far only four targets were reported showing
intervening absorption systems ($z>0$), all with low ion column
densities. \citet{fan05b} confirmed their first detection \citep{fan02} of an
absorption system at $z \sim 0.055$, along the sight line towards
PKS~2155-304. They found the observed \ion{O}{8} column density is
$\sim 4 \times 10^{15}$ \cms, and set a 3$\sigma$ upper limit 
on the \ion{O}{7}
column density of $\sim 10^{15}$ \cms at the same redshift. 
\citet{nic05} reported the detection of two absorption systems along the sight line
towards Mkn~421, but both systems showed \ion{O}{7} absorption lines
with column densities less than $10^{15}$ \cms. \citet{mat02}
reported a number of X-ray absorption systems towards H~1821+643 at 2
- 3 $\sigma$ level, and \citet{mck03} reported the detection
of an \ion{O}{8} absorption line along the sight line towards 3C~120.

To understand the difference between the local and intervening
absorption systems, we conduct a survey of a number of extragalactic
targets observed with {\sl Chandra} and {\sl XMM}-Newton to search for
local X-ray absorption lines. In this paper, we report the result
of this survey. We find that a model in which local X-ray absorbers
are associated with the Milky Way can explain our survey results
better than a model in which they are associated with the intragroup
medium in the Local Group. Recently, \citet{mck04} conducted a
systematic survey of 15 nearby AGNs to investigate the hot X-ray
absorbing gas in the vicinity of the Galaxy. While their work mainly
focuses on associating individual line of sight with known local
structures, our work is different in that we study the  generic
properties of this hot gas. 

\section{Sample Selection and Data Reduction}

All the background sources are selected from the {\sl Chandra} and {\sl
XMM}-Newton data archives. We focus on instruments that can provide
both high spectral resolution and moderate collecting area in the
soft X-ray band, and this results in four different instrument
combinations: (1) RGS, the Reflection Grating Spectrometer; (2)
LETG (the Low Energy Transmission Grating) + HRC (the High Resolution
Camera); (3) LETG + ACIS (the Advanced CCD Imaging Spectrometer); and (4)
HETG (the High Energy Transmission Grating) + ACIS. The first one
(RGS) is on board {\sl XMM}-Newton and the last three are instruments
on board {\sl Chandra}\footnote{For instruments on board {\sl
Chandra}, see {\sl Chandra} Proposers' Observatory Guide under
http://cxc.harvard.edu/proposer/POG/html/. For instruments on
board{\sl XMM}-Newton, see {\sl XMM}-Newton Users' Handbook under
http://xmm.vilspa.esa.es/}. The instrumental resolving power is
typically characterized by the line response function (LRF), the
underlying probability distribution of a monochromatic
source. HETG-ACIS has the highest resolving power: the LRF of the
Medium Energy Grating (MEG) has a full width at half maximum (FWHM) of
$\Delta\lambda \sim 0.02$ \AA\, around 20 \AA.\, The other three
combinations have a roughly constant resolution of $\Delta\lambda
\approx 0.05\,\AA$ across the relevant wavelength range. In some
cases, a target has been observed with several different instrument
configurations, and we select those that give the highest 
number of continuum counts.

So far, nearly all the sources with local X-ray absorption lines show the
detection of \ion{O}{7} with high confidence, so in this paper we
concentrate on the \ion{O}{7} He$\alpha$ resonance line with a rest
wavelength of 21.6 \AA\,. To ensure the significance of the absorption
line, we select targets that have a strong continuum between 21 and 23
\AA\,. Specifically, we bin the spectrum to roughly half of the LRF
FWHM and select targets which have at least 10 counts per bin around
21.6 \AA\,. In this way, we can ensure a signal-to-noise ratio (SNR)
of at least 4 within the LRF. We test with other binning size and
find essentially similar results. 

\vbox{
\begin{center}
\footnotesize
\begin{tabular}{llcrrlrc}
\multicolumn{7}{c}{~~~~~~~ Table 1: \ion{O}{7} He$\alpha$ Absorption Line ~~~~~~} \\ \hline \hline
Target & Redshift & Inst.$^a$ & Cont.$^b$ & $\ew$$^c$  & $S/N$$^d$ & $\ewt$$^e$
& Note$^f$\\ 
\hline 
MS 0737+7441	& 0.315   & 1  & 10 & ...  & ... & 24 &\\ 
NGC 3227	& 0.0039  & 2  & 12 & ...  & ... & 22 & \\ 
NGC 4258	& 0.0015  & 2  & 17 & ...  & ... & 18 & \\ 
Ton S180	& 0.0620  & 3  & 17 & ...  & ... & 18 &\\ 	  
MCG 6-30-15	& 0.0078  & 4  & 23 & $18_{-8}^{+10}$ & 4.1 & 13 &\\ 
NGC 7469	& 0.0163  & 1  & 32 & ...  & ... & 13 & \\ 
NGC 4593	& 0.009	  & 2  & 34 & $17\pm9$ & 3.3 & 13 & 1 \\ 
Mkn 766		& 0.0129  & 1  & 35 & ...  & ... & 13 & \\ 
H 1426+428	& 0.129	  & 1  & 35 & ...  & ... & 13 & \\ 
Ton 1388	& 0.1765  & 2  & 37 & ...  & ... & 12 & \\	 
PKS 0558-504	& 0.137	  & 1  & 40 & ...  & ... & 12 & \\ 
Mkn 501		& 0.0337  & 1  & 40 & ...  & ... & 12 & \\ 
NGC 3783	& 0.0097  & 4  & 47 & $36^{+27}_{-11}$ & 6.3 & 9 &  2 \\
1H 1219+301	& 0.182	  & 1  & 50 & ...  & ... & 11 &\\ 
H 1821+643	& 0.297	  & 3  & 59 & ...  & ... & 10 &\\ 
3C 273		& 0.1583  & 3  & 70 & $28_{-6}^{+12}$ & 6.4 & 9 & 3 \\
NGC 5548	& 0.0172  & 1  & 140 & $10\pm5$ & 3.3 &  6 &4\\ 
NGC 4051	& 0.0023  & 1  & 144 & $17_{-6}^{+2}$ & 5.5 & 6 & 5 \\ 
PKS 2155-304	& 0.156	  & 3  & 350 & $15_{-3}^{+4}$ & 7.1 & 4 & 6\\  
Mkn 421		& 0.03    & 1  & 1500 & $9.4\pm1.1$ & 8.8 & 2 & 7 \\ 
\hline
\end{tabular}
\parbox{3.5in}{
\vspace{0.1in}  a. Instrument: 1 --- RGS; 2 --- LETG-HRC; 3 ---
LETG-ACIS; 4 --- HETG-ACIS.\\   b. Continuum level, in units of counts
per bin. The bin size is 0.02 \AA\,for HETG and 0.025 \AA\, for
others.\\ c. Equivalent width of detected \ion{O}{7} He$\alpha$
absorption line, in units of m\AA.\\ d. Signal-to-noise ratio. \\ e: the minimum
equivalent width that can be detected at 3$\sigma$. \\f. References on
previously reported detections: 1 -- \citet{ste03a}; 2 --
\citet{kas02}; 3 -- \citet{fan03}; 4 -- \citet{ste03b}; 5 --
\citet{pou04}; 6 -- \citet{fan05b}; 7 -- \citet{wil05}}
\end{center}
}

Both {\sl Chandra} and {\sl XMM}-Newton data are analyzed using
standard software, i.e., CIAO 3.1\footnote{See
http://asc.harvard.edu.} and SAS
6.0\footnote{http://xmm.vilspa.esa.es.}. We refer readers to
\citet{fan05a} for detailed data analysis procedures. The local \ion{O}{7}
He$\alpha$ absorption lines from 8 of our 20 targets targets have been
reported previously by other papers; however, to ensure consistency in
our data analysis, we re-analyze all the data sets and fit the
absorption lines to obtain the equivalent width (EW) independently
(see column 5 of Table~1), and we refer the readers to the references
we list in the column 7 of Table~1 for spectra of detected \ion{O}{7}
lines. Due to the complex nature of the sources in our sample and
possible residual calibration effects, instead of fitting the
continuum with a physically meaningful model such as an absorbed power
law, we fit the continuum between 21 and 23 \AA\, with a polynomial
of an order between 3 to 5, to effectively remove residual feature
on scales $\gtrsim 0.4$ \AA. The residual is
then fitted with a Gaussian at $\sim 21.6$ \AA, using ISIS
(Interactive Spectral Interpretation System, see \citealp{hde00})
\footnote{see http://space.mit.edu/ASC/ISIS/}. While in all cases we
confirm the detections (and non-detections) that were reported by the
original authors, in some cases our measured EWs are smaller than
those originally reported. For example, we obtain an EW of $17\pm9$
m\AA\ for NGC~4593, while \citet{ste03a} report a much larger value,
$45\pm31$ m\AA.\,A careful reexamination of both data analysis
procedures indicates that the main discrepancy comes from the
determination of the continuum level (Steenbrugge, private
communication.) Nevertheless, our method is more stringent and the
measured EWs can be taken as conservative lower  limits.

\begin{figurehere}
\psfig{file=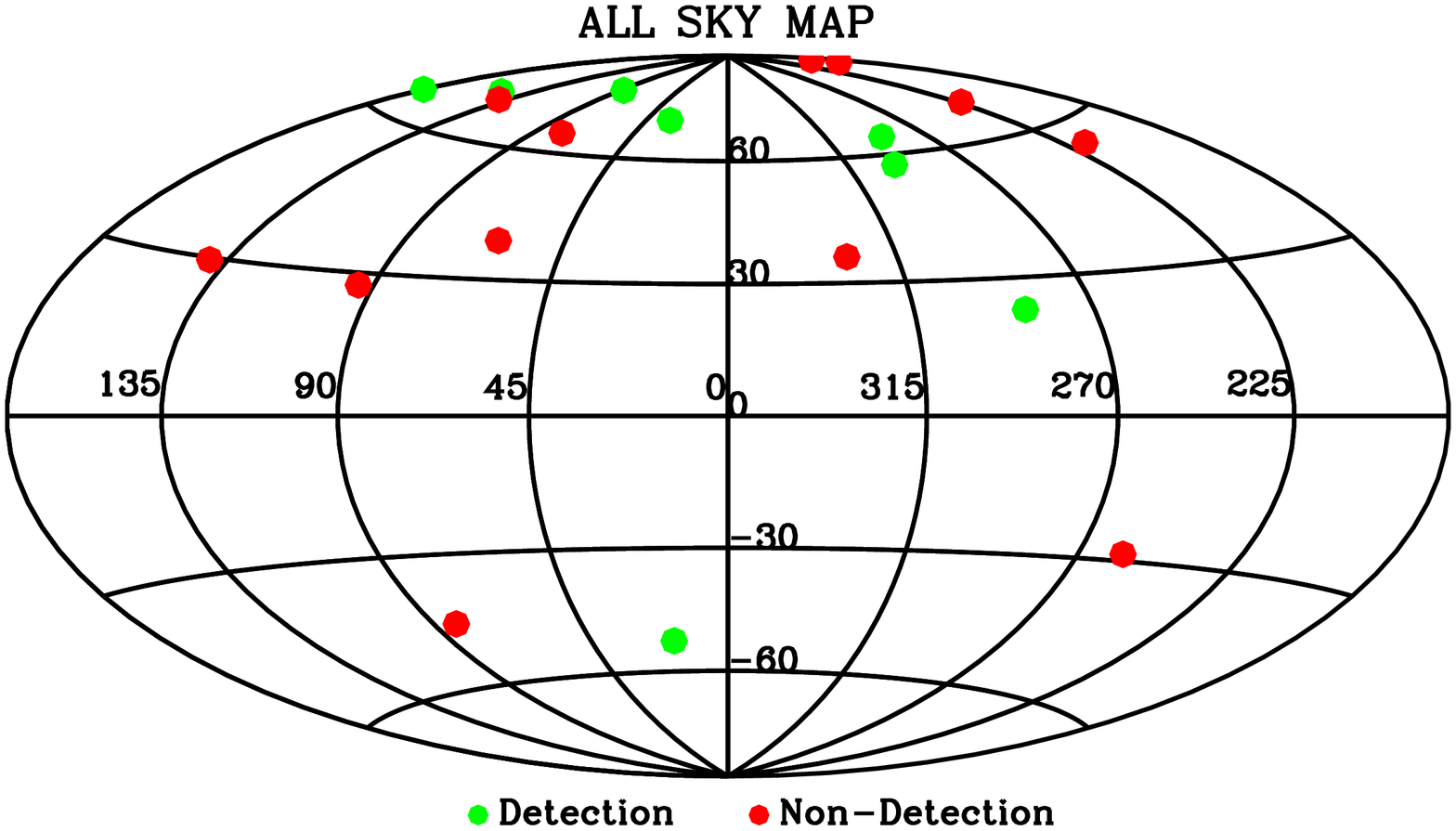,width=0.45\textwidth,angle=90}
\caption[h]{An all-sky Hammer-Aitoff projection of the 20 sight lines
  in our entire sample.}
\label{f1}
\end{figurehere}

Some of the sources in our sample are known to have intrinsic warm
absorbers. One may worry that the absorption lines detected at 21.6
\AA\ could be associated with the intrinsic warm absorbers,
particularly in the case of NGC~3783 where \ion{Ca}{16} may account
for part of the absorption \citep{kro03}. However, this does not have
a significant impact on our conclusions. Our case-by-case study of all
the warm-absorbers in our sample indicates that \ion{O}{7} and
\ion{Ca}{16} are the only possible sources of confusion. \ion{O}{7} is
unlikely because except in NGC~4051, none of the known warm-absorbers
show the outflow velocities that can compensate their redshifts. For
\ion{Ca}{16}, our study indicates that \ion{O}{8} always co-exists
with \ion{Ca}{16} but with at least $\sim 300$ times higher column
density, so a non-detection of \ion{O}{8} or a detection of \ion{O}{8} with column
density lower than $\sim 10^{18}$ \cms would simply rule out the
presence of \ion{Ca}{16}. By looking at each case, we find that
\ion{Ca}{16} can have a contribution only for NGC~3783 and
MCG--6-30-15. While \ion{Ca}{16} can contribute at most part of the
21.6 \aa line in NGC~3783, the contribution to MCG--6-30-15 is unknown
yet. To be conservative, we therefore consider a subsample that
excludes NGC~3783 and MCG--6-30-15.

The sight line to 3C~273 extends into the Galactic halo through the
edges of Radio Loops I and IV, which have been attributed to supernova
remnants (see, e.g., \citealp{egg95}). \citet{fan03} estimated that
the contribution to the \ion{O}{7} column density from the supernova
remnants accounts for at most 50\% of the observed equivalent length. 

If the absorption line is unsaturated, the equivalent width of
the \ion{O}{7} He$\alpha$ absorption line can be converted to a column
density by \citep{spi78}
\begin{equation}
N(\rm O VII) = 7 \times 10^{15}
\left(\frac{\ew}{20\rm\,mA}\right)\,cm^{-2} .
\label{eq:N}
\end{equation} 
However, saturation could be an important issue, as
revealed by high-order transition \ion{O}{7} lines discovered in the
Mkn~421 spectrum \citep{wil05}, the highest quality 
spectrum
 in our sample. In that case, the column density of \ion{O}{7},
 determined by high-order transition lines is $\sim 1.6\times10^{16}$
 \cms, more than five times higher than that estimated from
 Eq.(\ref{eq:N}). With this consideration, column densities converted
 from Eq.(~\ref{eq:N}) can serve as conservative lower
 limits, and we adopt $10^{16}\rm\,cm^{-2}$ as a fiducial value
in the following discussion
 for simplicity.

In Figure~\ref{f1} we show an all-sky projection of the entire
sample in our survey. The green dots show the positions of those
targets with detections, and the red dots show the positions of
those without detections.

We wish to estimate the fraction of the sky covered
by gas with an \ion{O}{7} equivalent width of at least EW,
$C(\ew)$. The difficulty we face is that we have a limited
sample size, and not all sources in our sample have
enough counts to permit detection of an absorption line
of equivalent width EW. For each source, we define the
threshold equivalent width, $\ewt$, which is the minimum
equivalent width that can be detected at 3$\sigma$:
\begin{equation}
\ewt=3\left(\frac{\lambda}{F_\lambda ATR}\right)^{1/2},
\end{equation}
where $\lambda$ is wavelength, $F_\lambda$ is the photon flux, $A$ is the area of
the detector, $T$ is the
observation time, and $R\equiv \lambda/\Delta\lambda$ is the
resolving power where $\Delta\lambda$ is the width of a resolution
element. $\ewt$ therefore depends on both
the detector and the source; 
as more data are gathered for a given source, its
value of $\ewt$ will decrease. 
Table 1 includes values of $\ewt$ for each source in
our sample.

\begin{figure}
\centerline{\psfig{file=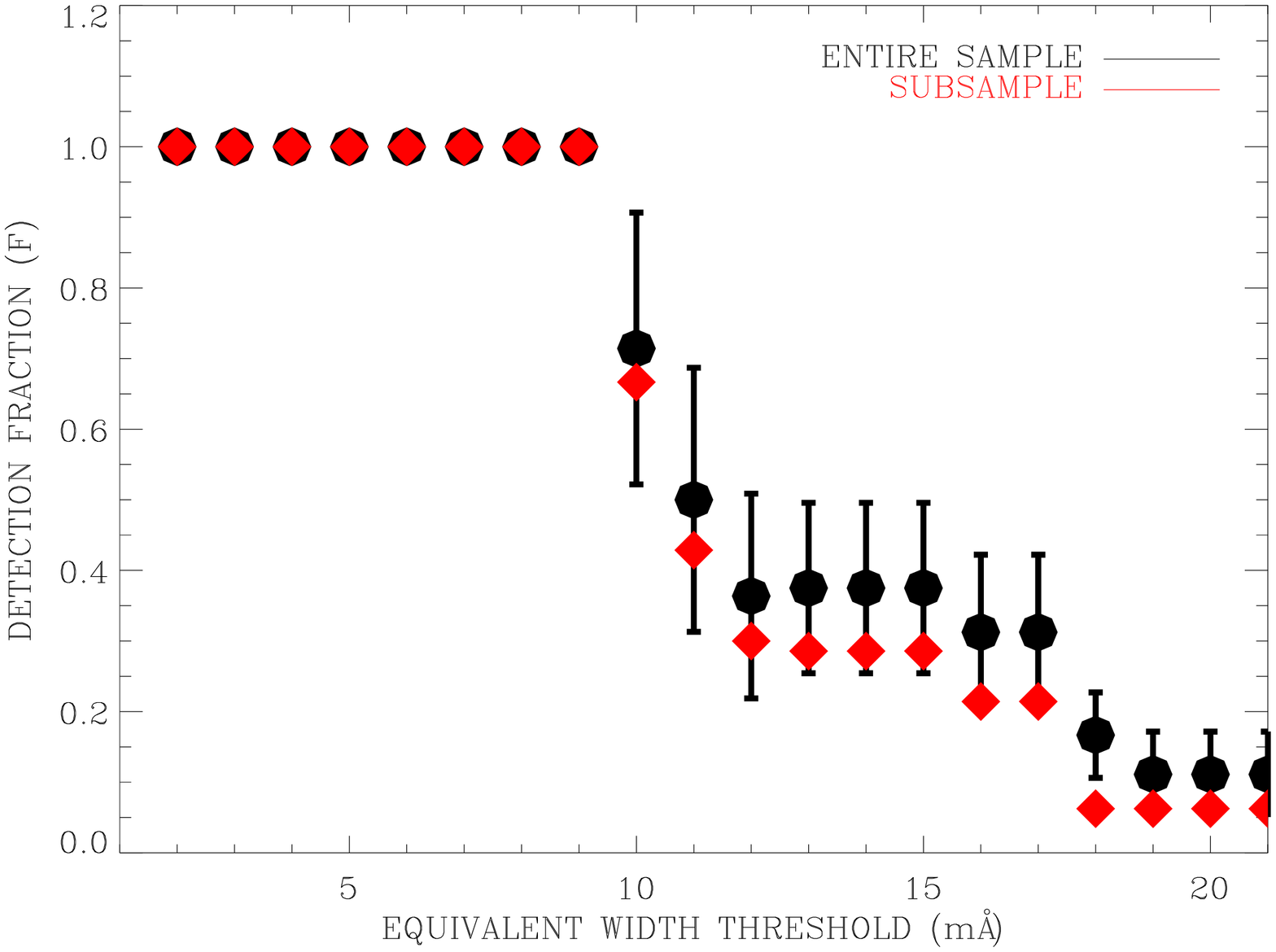,width=0.45\textwidth,angle=90}}
\caption[h]{The covering factor as a function of the equivalent width
  threshold. Dark line shows the entire sample and red line shows the
  subsample, and the error bars are the standard 1$\sigma$ error for binomial distribution.}
\label{f2}
\end{figure}

    To estimate $C(\ew)$, we consider all the sources
with $\ewt<\ew$, since only these sources have enough
counts to ensure detection of an absorption system with
an equivalent width $\ew$.
Let $\caln(\ew)$ be the number of these sources, and let 
$\caln_{\rm det}(\ew)$ be the number of sources
in this sample with detectable OVII absorption
(by definition, such sources must have $\ew\geq\ewt$).
The detection fraction is then 
\begin{equation}
F(\ew)\equiv\frac{\caln_{\rm det}(\ew)}{\caln(\ew)},
\end{equation}
and in the limit of large $\caln_{\rm det}(\ew)$ it will
approach $C(\ew)$. Figure~\ref{f2} shows the detection fraction as a
function of the minimum detectable equivalent width, and the error
bars are the standard 1$\sigma$ errors for a binomial distribution. We
show the error bars for the entire sample only for demonstration purposes. The
detection fraction starts at $\sim$10\% for the entire
sample and $\sim$ 5\% for the subsample, corresponding to lines with
${\rm EW} \gtrsim 20$~m\AA, and then gradually increases to 100\% with
increasing continuum level or decreasing EW threshold. The most
striking feature is that the detection fraction approaches 100\% as the
sensitivity approaches $\ew \sim 10$ m\AA, which occurs at a continuum
level of $\sim 60$ counts per bin. While it is tempting to infer that
the detection fraction would be 100\% if our sample had enough
sensitivity, we must be cautious: given the detections of \ion{O}{7}
absorption lines in the five highest-quality data sets, we can
conclude only that we have 90\% confidence that the sky detection
fraction is at least $(1-0.9)^{1/5}$, or 63\%.
 
\section{Where is the X-ray absorbing gas?}

While both {\sl Chandra} and {\sl XMM}-Newton have unprecedented
resolving power, the highest resolution we can achieve (HETG-ACIS, in
this case) is about $300$ \kms. Based on the Hubble flow and adopting
a Hubble constant of $H_0 = 70\rm\,km\,s^{-1}Mpc^{-1}$, this
corresponds to a distance
of $\sim 4$ Mpc at $z \sim 0$, which makes it impossible to
distinguish between a Milky Way-origin (with a halo radius of $\la
0.1$ Mpc) and a Local Group-origin (with a radius of $\sim 1$ Mpc) for
the X-ray absorbers. Nevertheless, with such a large sample size, we
can begin to understand the properties of these absorbers, i.e., their
size, density, temperature, 
spatial
distribution, etc. In the previous section, we estimate the sky
detection fraction $F$ of this hot gas. The detection fraction $F$ can
be taken as an estimate of sky covering fraction, defined as $C \sim
F$. Our estimation then indicates that the sky cover fraction is at
least higher than 63\%, and is likely close to unity. In the
following, based on this estimation we (1) calculate the expected
number of absorbers along lines-of-sight toward distant AGNs, (2) make
joint analysis with X-ray emission measurements, and (3) estimate the
total baryonic mass. Our calculation indicates that the observed X-ray
absorbers are associated with our Galaxy.                 

\subsection{Expected Number of Absorbers along LOSs toward Distant AGNs}

Given the high quality of spectra in our sample, any absorbers
distributed between us and background AGNs with similar column
densities, i.e., $\rm N(O{\rm VII}) \sim 10^{16}$ \cms, would be
detected with high confidence. Now let us estimate the expected
number of {\it intervening} absorbers with a similar \ion{O}{7}
column density in our sample.       

A key assumption in our model is that the X-ray absorbers detected at
$z \sim 0$ are not unique, i.e., an observer located in another
galaxy similar to the Milky Way should be able to see similar X-ray
absorption. The next step is to convert the covering fraction that we
presented in the previous section to the detectability of these
absorbers around systems similar to the Milky Way or the Local Group,
along the sight lines towards background AGNs. 

We assume that the X-ray absorbers are
uniformly distributed within the halo (of
a galaxy or a group of galaxies) that has a radius $R$.
Let $C$ be the sky covering factor as observed from
the center of the distribution of absorbers, which can be
obtained from observation.
If we view the absorbers in a different
group of galaxies, the observed covering factor would be $\frac 43 C$ (see, e.g., \citealp{tum05}).
Given the spatial density of the absorbers
within the halo, we can then calculate the expected number of absorbers
along the total line of sight towards all the background sources in
our sample. If the spatial
density of halos with radius $R$ is $\phi_0$, the expected number is then
\begin{equation}
\caln_{\rm obs} =\int_0^{\ell_{\rm tot}} \frac{4C}{3}\phi_0\sigma_0 d\ell,
\end{equation} 
where $\sigma_0 = \pi R^2$ is the cross section and
$\ell_{\rm tot}$ is the cumulative
distance along the sight lines to all the targets in the sample. 

\begin{figure}
\centerline{\psfig{file=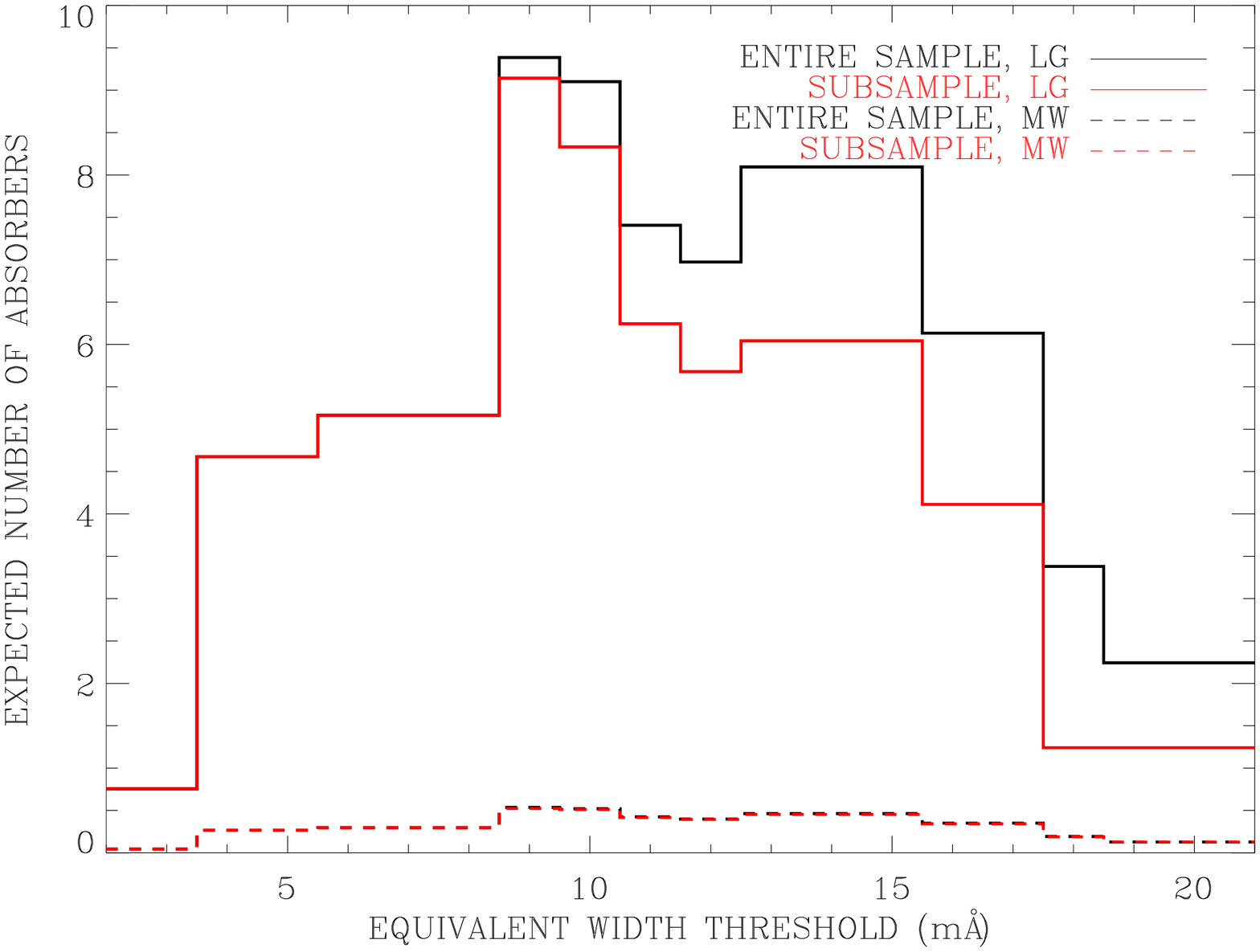,width=0.45\textwidth,angle=90}}
\caption[h]{The expected number of absorbers as a function of the
  equivalent width threshold. Local group case and the Milky Way case are
represented by solid and dashed lines, respectively. Dark lines are
for the entire sample, and red lines are for the subsample.}
\label{f3}
\end{figure}

In reality, both the covering fraction and the available pathlength,
and so $N_{obs}$, depend on the detection threshold $\ewt$. Taking
the covering fraction from Eq.~(3), we can calculate $N_{obs}$ from
Eq.~(4), as a function of $\ewt$ (Figure~\ref{f3}). Let us consider two scenarios,
in which the absorbers can be associated either with Milky Way-type
halos or with Local Group-type halos. For the Milky Way-type halo, $R
\la 100$ kpc, and the spatial density of halos is just the spatial
density of $L_{\star}$ galaxies, i.e., $\phi_0 \sim
0.004\rm\,Mpc^{-3}$ \citep{bla03}. On the other hand, if the observed
local absorbers are associated with the Local Group, $R$ would be
$\sim 1$ Mpc. An integral over the Press-Schechter mass function
\citep{pre74} from $10^{12}$ to $10^{13}\rm\,M_{\odot}$ gives a halo
spatial density of $\phi_0 \approx 0.0007\rm\,Mpc^{-3}$. In
figure~\ref{f3}, Local group case and the Milky Way case are
represented by solid and dashed lines, respectively. Dark lines are
for the entire sample, and red lines are for the subsample. Given the
fact that {\it none of the targets in the sample show any intervening
absorption \ion{O}{7} with column densities even close to $5\times
10^{15}$ \cms}, clearly the observations are consistent with the
association of the local X-ray absorbers with Milky Way-type halos,
but inconsistent with a Local Group association. In the Local Group
case, for the entire sample the maximum expected number of absorbers
is $\sim9.4$ if we take a threshold of $\sim 9$ m\AA.\ The Poisson
probability of detecting zero if the expectation is 9.4 is just
$\exp(-9.4) \sim 0.008\%$, at above Gaussian $3.7\sigma$ level. If we take
the subsample, the expected number decrease to $\sim 9.1$, and the
probability increases to $\sim 0.01\%$, roughly corresponding to a
Gaussian $3.7\sigma$ level.

\subsection{Mass Estimation}

We can also estimate the total baryon mass in these X-ray
absorbers. 
Let $n_{\rm H}$ be the hydrogen density
in an absorber and let $f_a$ be the volume filling factor
of the absorbers in the halo. 
Let $N_{\rm H}$ be the typical hydrogen column density along
lines of sight in which X-ray absorption is seen. Averaged
over the sky, the average column density is $CN_{\rm H}
=f_a n_{\rm H}R$, 
so that the baryonic mass of the absorbers is 
$M_b = (4\pi R^3/3)f_a \mu_{\rm H}n_{\rm H}=
(4\pi R^2/3)C\mu_{\rm H}N_{\rm H}$,
where $\mu_{\rm H}=2.3\times 10^{-24}$~g is the mass per H. 
Let $f$(\ion{O}{7}) be the ionization fraction of \ion{O}{7}\
and $Z/Z_\odot$ be the abundance of oxygen relative to
the solar abundance (taken as $4.6\times 10^{-4}$ --- \citealp{asp04}). We then have
\begin{equation}
M_b=\frac{4\pi}{3}C R^2 \mu_{\rm H}\left[
\frac{N({\rm O\ VII})}{4.6\times 10^{-4}f({\rm O\ VII})Z/Z_\odot}\right]\; ,
\end{equation}
where the factor in brackets equals $N_{\rm H}$.
If the absorbers
were distributed throughout 
the Local Group so that $R\sim 1$~Mpc, their total mass 
would be $M_b \simeq 
1\times 10^{12}C/[(Z/Z_\odot)f$(\ion{O}{7})]~$M_{\odot}$.
Since the dynamical mass of the Local Group is $\sim 2\times
10^{12}\ M_{\odot}$ (see, e.g., \citealp{cou99}), the total
baryon mass in the Local Group is $\sim 3\times 10^{11}\
M_{\odot}$ for a baryon fraction of $\sim$ 15\%. This means the covering fraction at most can be 30\% for solar metallicity. Since figure~\ref{f2} clearly indicates that the covering fraction can be much higher (and probably be as high as $\sim 100\%$) given enough instrumental sensitivity, we conclude that a Local Group origin for these X-ray absorbers is
unlikely because their estimated total mass would exceed
the baryon mass of the Local Group.
Our conclusion is consistent
with that of \citet{col05}
On the other hand, if we
associate these absorbers with our Galaxy, then the radius decreases
by a factor of 10-100 and the total mass 
is substantially reduced. 

\subsection{Joint Analysis with X-ray Emission Measurements}

Having presented evidence associating the $z=0$ X-ray absorbers with
the hot gas within the Milky Way, we can further constrain the
properties of the X-ray absorbers by assuming that the same absorbers
are also responsible for at least some of
the hot halo foreground observed in X-ray
emission. The soft X-ray background is believed to be produced by three
components: the Local Hot Bubble (LHB), the extragalactic background
(mainly from point sources), and a halo component. 
By analyzing {\sl Rosat} All Sky Survey (RASS) data, \citet{kun00} concluded that the halo emission actually
consists of two components, a hard component with $\log T_H = 6.46$
and a soft component with $\log T_S = 6.06$. 
Because the hard
component is too hot to produce a substantial amount of \ion{O}{7} in
collisional ionization equilibrium, we assume that most of the observed
\ion{O}{7} resides in the soft component.
 
Let $\emo$ be the observed emission measure.  Since $\emo$ is based on
solar abundances and is averaged over the sky,  it is related to the
actual emission measure, EM, along a line of sight through an absorber
by $\emo\simeq (Z/Z_\odot)C\; {\rm EM}=(Z/Z_\odot)C   n_e n_p \lpc$,
where $n_e\simeq 1.2 n_p$ and $n_e$ and $n_p$ are the electron and
proton densities. $\lpc$ is the average pathlength through an absorber
in pc. \citet{kun00} found an emission measure of $\emo = 4.7 \times
10^{-3}\rm\,cm^{-6}\, pc$  for the soft component.  From the
absorption measurement, we have  $N({\rm O\; VII})=4.6\times 10^{-4}
(Z/Z_{\odot})f_{OVII}n_{\rm H}L \sim 10^{16}$ \cms. Since
$f_{OVII}\simeq 1$  at $\log T_S = 6.06$, we have $n_{\rm H} \simeq
5.6\times 10^{-4}/C  \rm\; cm^{-3}$ (independent of metallicity), and
$L \simeq 13C(Z_\odot/Z)$ kpc.  This also provides independent
evidence that these X-ray absorbers should be associated with our
Milky Way instead of the intragroup medium in the Local Group.

\section{Discussion}

\citet{ras03} argued that based on a combination of \ion{O}{7} emission line
and absorption line measurement, the scale length of the O VII
absorber should be at least $140/(Z/Z_{\odot}/0.3)$ kpc. However,
their argument depended crucially on the temperature of the
emitter/absorber. A shift in temperature by 0.1 to 0.2 in log
space will change the scale length by a factor of 10. They
determined the temperature should be between $2 - 5\times 10^6$ K,
based on the 
\ion{O}{8} and \ion{O}{7} line ratio. A recent work, based on much
higher quality Chandra data on Mkn 421 \citep{wil05}, indicated that
just from Chandra measurement of the \ion{O}{8} and \ion{O}{7} line ratio,
the temperature of the Mkn 421 local absorber must be lower than
$1.6\times10^6$ K. Such a temperature brings the scale
length down to the Galactic scale.

We estimate this X-ray absorbing gas has a density of a few $\times
10^{-4}\rm\; cm^{-3}$ with temperature around $10^6$ K. Our
estimations are consistent with values predicted from other models
such as the dynamics of the Magellanic Stream \citep{moo94}. \citet{hec02}
suggested, based on observations of \ion{O}{6} absorption in the disk
and halo of the Milky Way and in the intergalactic medium, that these
absorption systems belong to radiative cooling flow of hot gas. Their
predicted \ion{O}{7} column density is consistent with what we find in
this paper. A key question then is: how to keep this gas hot without
cooling? For gas with such a high density, the cooling time will be
less than the Hubble time. In collisional ionization equilibrium, the
typical cooling time scale is
\begin{equation}
t_c \sim 0.25\; \frac{T_6}{n_{-3}\Lambda_{-22}}~~{\rm Gyr},
\end{equation} 
with $n_{-3} = n_H/(10^{-3}\,\rm
cm^{-3})$ and $T_6= T/(10^6\,\rm K)$. Here, $n_{\rm H}^2\Lambda_{-22}$ is the
radiative cooling rate in units of $\rm 10^{-22}\,erg\,cm^{-3}s^{-1}$.
Under the assumption that the cooling rate at $T_6\sim 1$ scales
as the Fe abundance, the cooling rate of \citet{sut93} is $\Lambda_{-22}\simeq 0.85 (Z/Z_\odot)T_6^{-0.8}$,
where the solar abundances are taken from \citet{asp04}. 
For the density and temperature we have estimated
for the absorbers, the cooling time is much less than
the Hubble time, unless the metallicity is extremely low ($Z/Z_\odot
\la 0.025$). 

To keep this relatively dense gas ($n_e \lesssim 10^{-3}$ \cmq) at
temperatures around $10^6$ K without cooling down over a Hubble time
scale, some sort of heating mechanism must play an important
role. Supernova heating is a potential candidate. The cooling rate,
according to the calculations above, would be
\begin{equation}
{\cal L} = n_{\rm H}\Lambda (M_b/\mu_{\rm H}) \approx
4.2\times 10^{41} (R/100\;{\rm kpc})^2~~~\rm erg\,s^{-1},
\end{equation} 
for $T_6\simeq 1$; note that the dependence on $C$, $f$(O~VII),
and $Z/Z_\odot$ has canceled out.
Assuming that the supernova rate is about 0.02 yr$^{-1}$
in our Galaxy (see, e.g., \citealp{dra99}), and that the energy output of each supernova explosion is
about $10^{51}\rm\,erg$, the heating rate would be 
about $ 6.3\times10^{41}\,\rm erg\,s^{-1}$. For
$R\ll 100$ kpc, which is consistent with our estimate
for $L$, only a small fraction of the supernova energy
is needed to maintain the temperature of the absorbers 
at $T\sim 10^6$~K.

In this paper, we find that the sky covering fraction of this hot,
\ion{O}{7}-absorbing gas is at least 63\%, at 90\% confidence. This is
based on the detections of \ion{O}{7} absorption lines at $>3\sigma$
level in the spectra of the top five high quality spectra. While we cannot
obtain the exact location of this hot gas within our Galaxy, joint
analysis with the X-ray background data indicates the scale height of
this hot gas should be $\simeq 13C(Z_\odot/Z)$ kpc. Finally we
conclude that, based on three independent estimations, the X-ray
absorbing gas detected locally are part of the hot gas in our
Galaxy. Future observations will provide robust test of this result.

\smallskip
{\it Acknowledgments:} We thank Julia Lee for help with MCG~6-30-15
data and Rik Williams for help with Mkn~421 data. We also thank
the anonymous referee for valuable suggestions. TF was supported
by the NASA through {\sl Chandra} Postdoctoral Fellowship Award Number
PF3-40030 issued by the {\sl Chandra} X-ray Observatory Center, which
is operated by the Smithsonian Astrophysical Observatory for and on
behalf of the NASA under contract NAS 8-39073. The research of CFM was
supported in part by NSF grant AST00-98365 and by the support of the
Miller Foundation for Basic Research. MGW was supported in part by a
NASA Long Term Space Astrophysics Grant NAG 5-9271.

\end{document}